\documentclass[prl,twocolumn,superscriptaddress,showpacs,floatfix]{revtex4}

\usepackage{amssymb}

\usepackage{afterpage}

\usepackage{epsfig}
\usepackage{color}

\begin{document}

\title{Onset of flow in a confined colloidal glass under an imposed shear stress}

\author{Pinaki Chaudhuri}
\affiliation{Institut f\"ur Theoretische Physik II, Heinrich-Heine-Universit\"at D\"usseldorf, 
40225 D\"usseldorf, Germany}
\affiliation{Johannes-Gutenberg-Universit\"at Mainz, Institut f\"ur Physik, WA 331, 55099 Mainz, Germany}
\author {J\"urgen Horbach}
\affiliation{Institut f\"ur Theoretische Physik II, Heinrich-Heine-Universit\"at D\"usseldorf, 
40225 D\"usseldorf, Germany}

\begin{abstract}
A confined colloidal glass, under the imposition of a
uniform shear stress, is investigated using numerical simulations. Both
at macro- and micro-scales, the consequent dynamics during the onset
of flow is studied. When the imposed stress is gradually decreased,
the time-scale for the onset of steady flow diverges, associated with
long-lived spatial heterogeneities.  Near this yield-stress regime,
persistent creep in the form of shear-banded structures is observed.
\end{abstract}

\maketitle

A defining mechanical property of amorphous solids is the existence
of a yield stress; these materials yield only when the applied stress
exceeds this threshold \cite{bonn}.  In experiments, the material's
yielding response is commonly probed by using stress as a control
parameter \cite{larson}.  Such measurements report about existence of
a creeping regime, where the material deforms very slowly.  In soft
amorphous materials (e.g. colloids, gels, emulsions etc.), steady
flow is often not observed in such situations within experimental
durations \cite{bonn2, laurati}. This is related to recent recognition of
diverging time-scales for the onset of flow \cite{catherine,thomas,divoux}.
Beyond these macro-measurements, little is known about the spatio-temporal
characteristics of the local dynamics around yielding, which would allow
for improved understanding of the micro-mechanisms at play.

An emerging scenario for the micro-dynamics is that when the material is
sheared, structural rearrangements occur locally \cite{jl1,anael,stefan}
which subsequently trigger more events nearby and this correlated
process initiates and sustains the flow.  Kinetic elastoplastic
models \cite{bocquet} based on such a picture of fluidization have
been able to explain various steady-state flow properties in confined
systems \cite{goyon,jop,pinaki1}.  Further, it has been conjectured
that the triggering process results in an avalanche-like behaviour,
as evidenced via simulations \cite{lemaitre,smarajit}. Here, we focus
on such micro-macro processes when a quiescent glass is subjected
to an external stress.  Recent creep measurements in a carbopol gel
\cite{divoux} as well as in a granular mixture \cite{clement1, clement2}
indicate that in those materials flow occurs via the formation of local
plastic events which build up to form transient shear-bands.  However,
whether these flow inhomogeneities are generic to the fluidization of
amorphous systems under imposed stresses and linked to the correlated
processes discussed above, needs clarification.

In this Letter, we report a simulational study of a confined colloidal 
glass under an imposed {\it uniform shear stress} to elucidate the
nature of the local dynamics during the {\it onset of flow}. 
We use a geometry that mimics the typical planar Couette setup
in stress-controlled experiments of soft materials. We show that indeed
the time-scale for the onset of steady flow diverges when the stress
is decreased, along with the emergence of nonlinear creep (similar to
a range of materials \cite{divoux,thomas,andrade,paper2,rocks,paper}).
Moreover, by following the local dynamics, we observe that the mobility
can be spatio-temporally heterogeneous and the mobile regions take the
shape of shear-band-like structures. These structures have a lifetime
which increases with decreasing stress, which is linked to the delayed
onset of steady flow. Thus, our work provides evidence that in the
presence of an external shear stress, the fluidization of the glass
near yielding is extremely slow, along with the presence of persistent
spatial heterogeneities.

In our simulations,  we consider the model colloidal system of a $50:50$
binary Yukawa fluid (for details of interactions and model parameters,
see \cite{zausch, winter}).  Our molecular dynamics simulations
have been done for samples consisting of $N=12800$ particles, having
the dimensions $L_x=26.66\,d_s$, $L_y=53.32\,d_s$, $L_z=13.33\,d_s$
($d_s$ is the diameter of the smaller particles).  We work in the $NVT$
ensemble and the temperature control is done by using a Lowe thermostat
\cite{lowe}.  At a high temperature of $T=0.2$, we equilibrated the
system using periodic boundary conditions.  Then, $m=24$ independent
configurations sampled at $T=0.2$ were instantaneously quenched to
$T=0.05$ (below the mode-coupling critical temperature of $T_c=0.14$
\cite{zausch}).  Each of these $m$ configurations are then aged for
durations of $t_{\mathrm{age}}=10^3$, $10^4$, $10^5$.  When each
configuration reaches a certain $t_{\mathrm{age}}$, we freeze the
particles at $0<y<2\,d_s$ and $L_y-2\,d_s<y<L_y$ to prepare glassy states
confined between rough walls. Then, these confined samples are sheared
by pulling the top plate in $+{x}$ direction with a fixed force $F_0$
(similar to rheometers) \cite{varnikyield}, which imposes a constant
shear stress of $\sigma_0=F_0/[L_xL_z]$; we study the response of the
confined glass to such a field.

\begin{figure}
\includegraphics[scale=0.34,angle=0,clip=true]{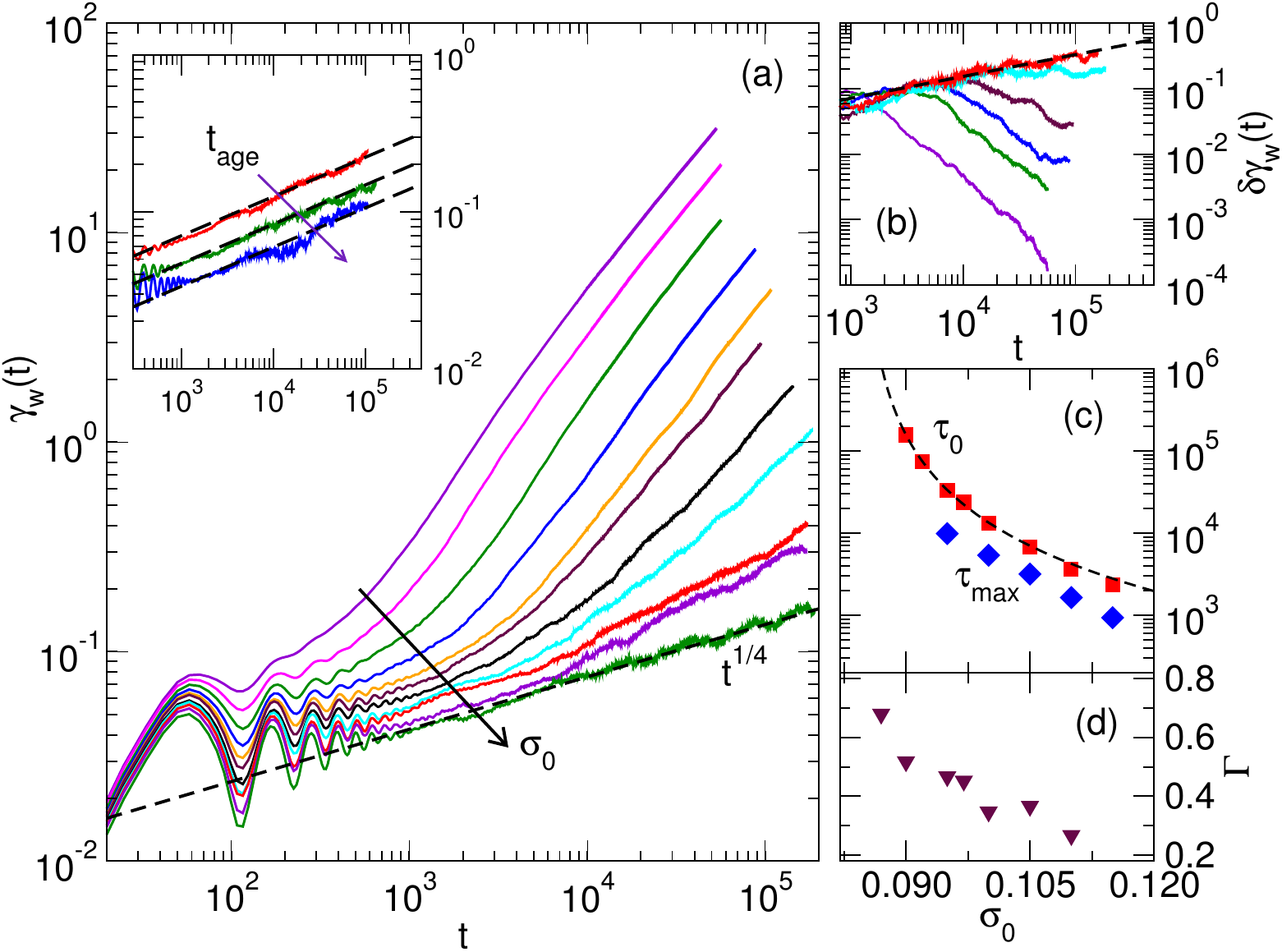}
\caption{(a) Evolution of strain at wall, $\gamma_{\mathrm w}(t)$
for a range of imposed stress, $\sigma_0 = 0.115$, $0.110$, $0.105$,
$0.100$, $0.097$, $0.095$, $0.092$, $0.090$, $0.087$, $0.085$, $0.080$
(from top to bottom), at $t_{\mathrm age}=10^4$. (Inset) $\gamma_{\mathrm
w}(t)$ at $\sigma_0=0.08$ at different ages ($t_{\mathrm{age}}=10^3$,
$10^4$, $10^5$), with dashed lines corresponding to power law fits
$\gamma_{\mathrm w}(t)\sim{t^{1/4}}$.  (b) Strain fluctuations across
trajectories ($\delta\gamma_{\mathrm w}(t)$) for different $\sigma_0$;
the dashed line being a fit with ${t^{1/3}}$.  (c) Variation with
$\sigma_0$: (In squares) Timescale ($\tau_{0}$) to reach $\gamma_{\mathrm
w}=1$; the dashed line is a fit with $A/(\sigma_0-\sigma_s)^{\beta}$
($\sigma_s=0.0848$, $\beta=2.285$).  (In diamonds) Timescale ($\tau_{\rm
max}$) at which the peak in $\delta\gamma_{\mathrm w}(t)$ occurs.
(d) With changing $\sigma_0$, fluctuations in timescales ($\Gamma$)
to reach strain of $0.3$.}
\label{fig1}
\end{figure}
Similar to experimental creep measurements, we study the response
of the glass by monitoring the strain experienced at the top wall
$\gamma_{\mathrm w}(t)$. This is done by recording, for each of the $m$
samples, the velocity of the wall $v_{\mathrm w}(t)$ as a function of
time, and then the average effective strain at the wall is obtained:
$\gamma_{\mathrm w}(t)=\langle\gamma_{\alpha}(t)\rangle_e$, where
$\gamma_{\alpha}(t)=\int_0^t\![v_{\rm w}(t)/(L_y-4\,d_s)]\mathrm{d}t$ is
the wall strain computed for $\alpha$-th sample and $\langle\rangle_e$ is
an average over the ensemble of $m$ trajectories.  In Fig.~\ref{fig1}(a),
we show how $\gamma_{\mathrm w}(t)$ evolves for a wide range of imposed
stress $\sigma_0$. These curves show several regimes - (i) in all cases,
at early times, the strain increases initially and then there is an
oscillatory part, corresponding to the regime when the stress builds up
inside the confined sample - the oscillations occur due to the interplay
of the imposed stress at the wall and the restoring force of the deformed
glass; (ii) after this, for large values of $\sigma_0$, one quickly sees
an asymptotic linear regime in $\gamma_{\mathrm w}(t)$, which corresponds
to a steady flow at a fixed shear-rate; (iii) as $\sigma_0$ is decreased,
an intermediate regime in $\gamma_{\mathrm w}(t)$ emerges, which for
small enough $\sigma_0$ has a power-law behaviour - for the creep at
$\sigma_0=0.080$ (i.e.~below the dynamic yield stress $\sigma_d=0.0857$
\cite{hb}), we obtain a power-law exponent of  $0.25$. We also observe
that the underlying process responsible for creep is not dependent on
the age of the samples.  In the inset of Fig.~\ref{fig1}(a), we show
the strain data for the different ages ($t_{\mathrm{age}}=10^3$, $10^4$,
$10^5$): for all the three different ages, the exponent of the power law
remains the same.

For most values of $\sigma_0 > \sigma_d$, one eventually observes
the steady flow ($\gamma_{\mathrm w}(t)\sim{t}$).  One can define a
time-scale for onset of flow, $\tau_0$, as the time required for the
system to reach a strain of $\gamma_{\mathrm w}(\tau_0)=1$; using this
definition, we calculate $\tau_0$ for each $\sigma_0$ and plot it in
Fig.~\ref{fig1}(c) (squares). One can see that the time-scales increase
with decreasing stress and the data can be fitted with the function
$A/(\sigma_0-\sigma_s)^{\beta}$, with $\sigma_s=0.0848$ ($\approx
\sigma_d$, the estimated dynamical yield stress) and $\beta=2.285$
(which is similar to experimental observations \cite{divoux,thomas}).

In supercooled liquids, diverging time-scales are seen to be associated
with increasing heterogeneity in the dynamics \cite{ludo}. For the
onset of flow in glasses, we explore this possibility by studying the
fluctuations in response within the ensemble of $m$ samples having
the same age; this is quantified by calculating $\delta\gamma_{\mathrm
w}(t)=\langle{\gamma_{\alpha}^2(t)}\rangle_{e}/\langle{\gamma_{\alpha}(t)}\rangle^{2}_{e}-1$.
In Fig.~\ref{fig1}(b), we plot $\delta\gamma_{\mathrm w}(t)$ for
different $\sigma_0$.  We see that for large values of $\sigma_0$,
the function $\delta\gamma_{\mathrm w}(t)$ increases with time,
has a maximum and then decays at long times. The initial increase
in $\delta\gamma_{\mathrm w}(t)$ has a power-law behaviour, which
is identical for all $\sigma_0$. This is similar to what was earlier
observed in experiments of creep flow using paper samples \cite{paper}
-- a power-law in average strain, as well as a power law in the
fluctuations. We can also track the location of the peak ($\tau_{\rm
max}$) in $\delta\gamma_{\mathrm w}(t)$; the variation of $\tau_{\rm
max}$ with $\sigma_0$ is shown in Fig.~\ref{fig1}(c) - one can see that
the increasing trend is similar to that for $\tau_0$. $\tau_{\rm max}$
approximately corresponds to a macroscopic strain of $\gamma_{\mathrm
w}\sim{0.3}$ as well as the time-scale at which diffusion sets in. Thus,
the diverging time-scale for the onset of steady motion of the wall
is linked to the increasing delay in the onset of diffusion within
the system.

\begin{figure}
\centerline{\includegraphics[scale=0.35,angle=0,clip]{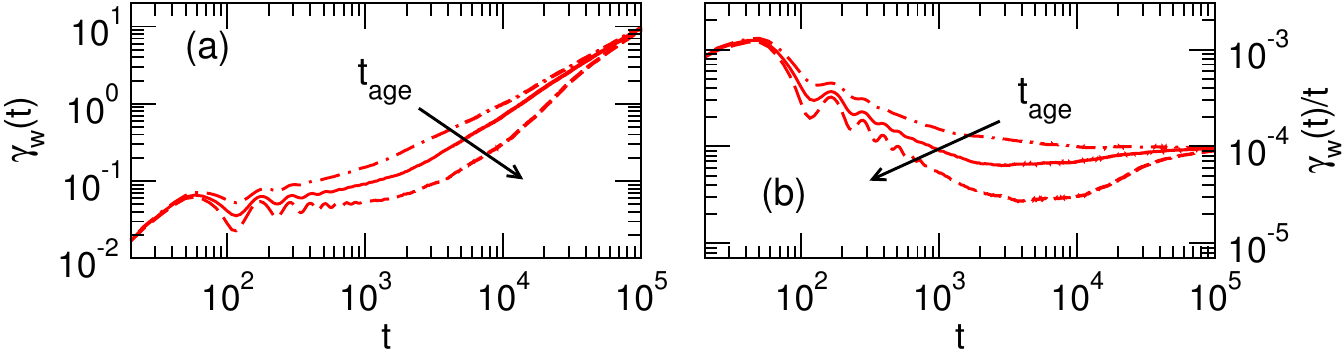}}
\centerline{\includegraphics[scale=0.5,angle=0,clip]{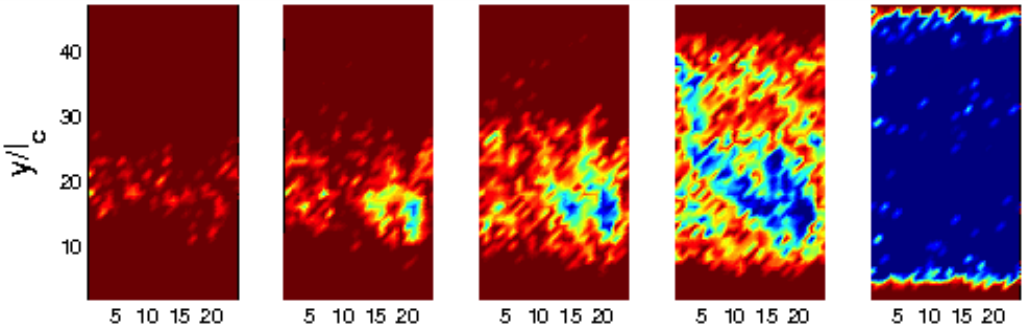}}
\centerline{\includegraphics[scale=0.5,angle=0,clip]{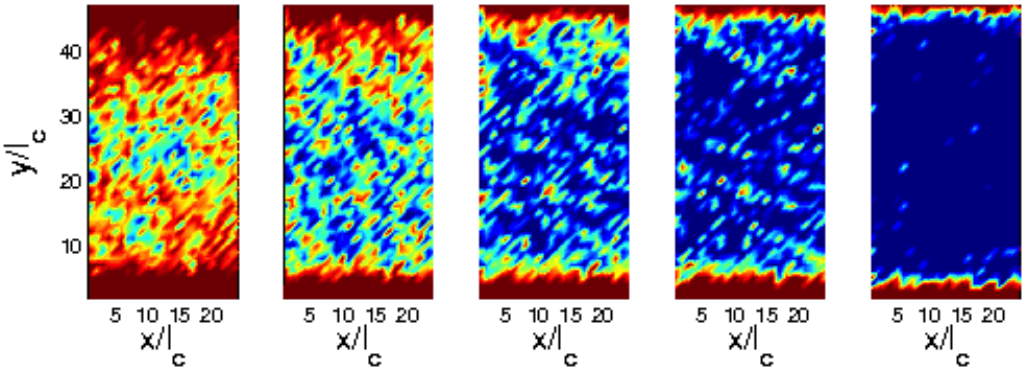}}
\centerline{\includegraphics[scale=0.51,angle=0,clip]{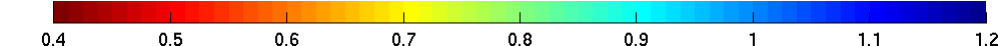}}
\caption{{\it Top panel}: (a) Wall-strain $\gamma_{\mathrm w}(t)$ for
samples having different ages ($t_{\mathrm{age}}=10^3$, $10^4$, $10^5$)
for imposed stress of $\sigma_0=0.1$. (b) Corresponding effective
strain-rates, $\gamma_{\mathrm w}(t)/t$ at the wall.  ({\it Middle panel})
Displacement maps for $t= 7470$, 14940, 22410, 29880, 85905 (from left to right)
at a trajectory from $t_{\mathrm{age}}=10^5$ ensemble.
({\it Bottom panel}) Maps at same times for a trajectory from $t_{\mathrm{age}}=10^3$ ensemble.}
\label{fig2}
\end{figure}

Till now, by monitoring the motion of the top wall, we have gauged
the macroscopic response of the material. We now want to  see how
deformation shapes up at a more local scale within the sample; whether
the local response is uniform or heterogeneous and whether any spatial
structures are formed as the system yields.  In order to study this,
we construct maps of transverse displacements (which are non-affine
motions occurring due to the local structural rearrangements) in the
following manner. Before we apply the external stress ($t=0$), we divide
the $xy$ plane of the simulation box into small square cells (of length
$l_c=1.1\,d_s$) and identify the particles in each cell. Next, after
time $t$, we calculate the transverse displacement of each particle
$\Delta{y}_i(t)=|y_i(t)-y_i(0)|$ and then calculate for each cell the
local mobility, $\mu_{lm}(t)=\langle\Delta{y}_i(t)\rangle_{lm}$, where
$\langle\rangle_{lm}$ is the average over all the particles in the cell
$\{lm\}$ at $t=0$, to construct the spatial maps.

We begin by using such maps to study the local dynamics for different
ages of the samples, at a fixed imposed stress.  For $\sigma_0=0.1$,
the evolution of the macroscopic strain $\gamma_{\mathrm w}(t)$ for
trajectories at $t_{\mathrm{age}}=10^3$, $10^4$, $10^5$ as well as
the corresponding effective strain-rates at the wall, $\gamma_{\mathrm
w}(t)/t$, are shown in Fig.~\ref{fig2}(a) and \ref{fig2}(b), respectively.
We see that the curvature in $\gamma_{\mathrm w}(t)/t$ changes with age
-- while the curve for $t_{\mathrm{age}}=10^5$ has a pronounced minimum,
it is a monotonically decreasing function for $t_{\mathrm{age}}=10^3$.
This implies that the intermediate flow states could vary for samples
having different ages. This we demonstrate via the mobility maps. The
time evolution of local mobilities for a trajectory belonging to the
ensemble with $t_{\mathrm{age}}=10^5$ is displayed in the middle panel
of Fig.~\ref{fig2} (with the time increasing from left to right). The
most important observation is that the transient local mobilities are
spatially heterogeneous. Initially, a few spots of mobility evolve to a
growing domain of large displacement. This then triggers mobility in the
neighbourhood to form a structure which looks like a shear-band (spanning
the length of the simulation box). This band then grows transversely
as the mobility spatially spreads and eventually the entire system
is fluidized.  One can contrast this with the spatial mobilities for a
trajectory from the ensemble with $t_{\mathrm{age}}=10^3$ (bottom panel
of Fig.~\ref{fig2}).  Unlike the aged sample, here the spatial spread
of large mobilities is quite uniform and it continues to be the case as
the entire system gets eventually fluidized.  Thus, the displacement
maps clearly reveal that the transient flow states differ with aging
\cite{shi}: for the more aged initial structure, an extended time window
exists over which shear-banding is observed before the system finally
becomes fully fluidized.  This is reflected in the aging dependence of
the compliance curves (see Fig.~\ref{fig2}) and should rationalize similar
experimental findings of Siebenb\"urger {\it et al.}~\cite{thomas}.

\begin{figure}
\includegraphics[scale=0.4,angle=0,clip=true]{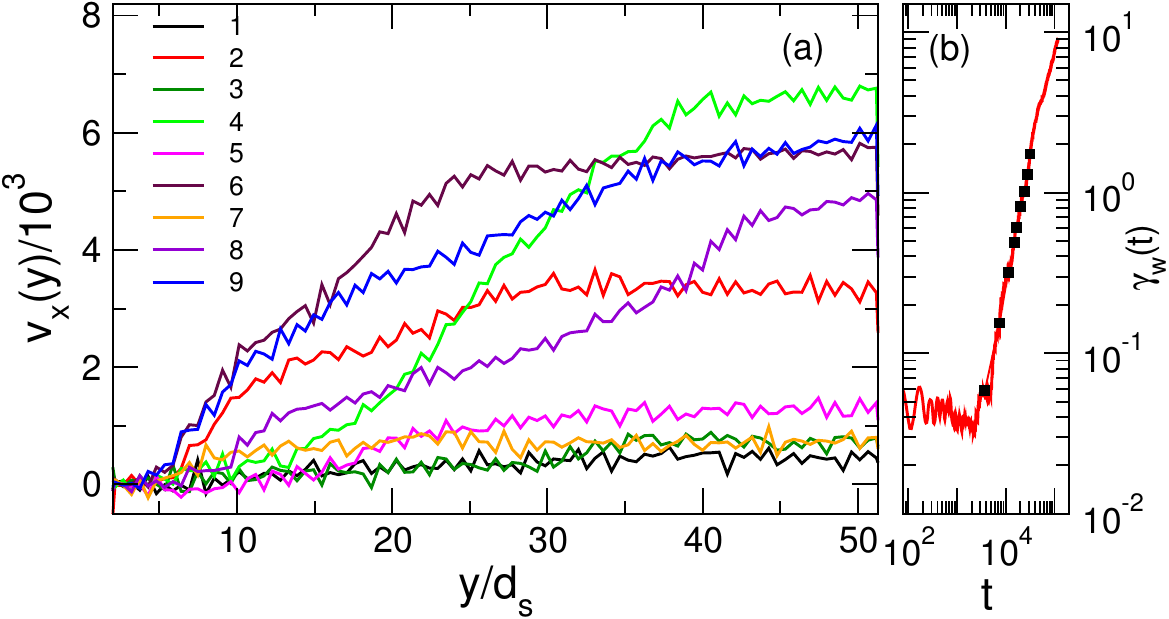}
\caption{(a) Velocity profiles during the onset of flow at
$\sigma_0=0.1$ for a sample with age $t_{\mathrm{age}}=10^5$, averaged
for $\Delta{t}=498$, starting from different time-origins $t_0$ (labelled
in sequence). (b) For the same trajectory, $\gamma_{\mathrm w}(t)$
with the black squares marking the different $t_0$.}
\label{fig3}
\end{figure}

\begin{figure*}
$\begin{array}{cc}
{\includegraphics[scale=0.88,angle=0,clip=true]{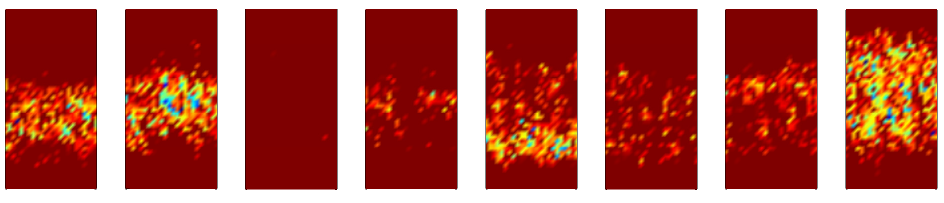}}&
{\includegraphics[scale=0.27,angle=0,clip=true]{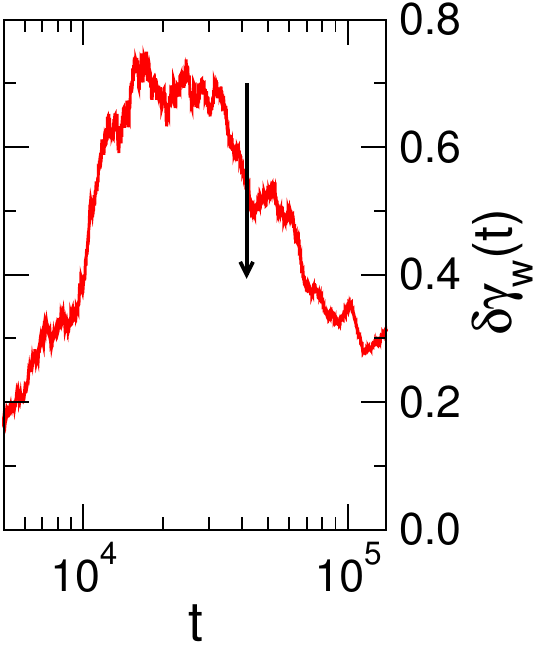}}
\end{array}$
\caption{For samples at $t_{\mathrm{age}}=10^5$ ({\it left to right}),
the first eight panels show displacement maps for different trajectories
at a fixed time $t=42745$ for $\sigma_0=0.092$, indicating the variation
in response of different samples at same $t_{age}$. Colorbar
same as Fig.\ref{fig2}. ({\it Last panel})
$\delta\gamma_{\mathrm w}(t)$ for the corresponding ensemble, with the
arrow indicating the time of measurement.}
\label{fig4}
\end{figure*}

Since for the aged samples the transverse displacement maps indicate
an inhomogeneous evolution of flow, one can expect that corresponding
velocity profiles (which monitor the spatial dependence of the motion
along the force direction) would also capture this behaviour.  At the
same imposed stress ($\sigma_0=0.1$), we show such velocity profiles
in Fig.~\ref{fig3}(a) for a trajectory sampled from the ensemble with
$t_{\mathrm{age}}=10^5$.  Since instantaneous profiles are too noisy, we
average velocity profiles over short intervals in time ($\Delta{t}=498$)
starting at different points in time ($t_0$) after the imposition of the
stress. In Fig.~\ref{fig3}(b), we plot the evolution of the wall-strain
$\gamma_{\mathrm w}(t)$ for this trajectory and the different $t_0$
are marked on the curve; they correspond to the time-scales when the
material ``breaks'' into flow.  We see that, during these times, the
velocity profiles fluctuate quite a lot, with signatures of intermittent
flow (i.e.~existence of little or no flow at some time instances,
interspersed with bursts of flow) as well as the existence of band-like
profiles (co-existence of regions of no-flow with more mobile regions).
These transient heterogeneities in the velocity profiles, while in
agreement with the observation in gels \cite{divoux}, reflect a more
complex motion in flow direction.

\begin{figure}
\includegraphics[scale=0.45,angle=0,clip]{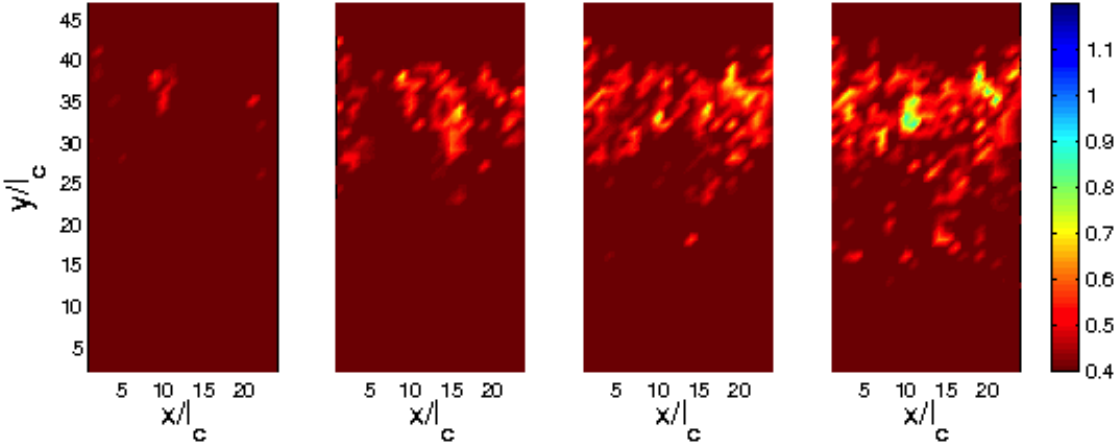}
\caption{For $\sigma=0.08$, displacement maps along a single trajectory
from the $t_{\mathrm{age}}=10^4$ ensemble. Measurements are done at $t=10790$, 25730, 92960,
251905 ({\it from left to right})}
\label{fig5}
\end{figure}

Next, mobility maps are used to clarify the origin of the
non-monotonicity in $\delta\gamma_{\mathrm w}(t)$. For samples aged
to $t_{\mathrm{age}}=10^5$, we study the dynamics for the independent
trajectories evolving for $\sigma_0=0.092$. In Fig.~\ref{fig4}(a), eight
maps, calculated at $t=42745$, from this ensemble are shown; the time
of measurement is marked by an arrow in Fig.~\ref{fig4}(b), which shows
$\delta\gamma_{\mathrm w}(t)$ for the entire ensemble.  From map-to-map,
one clearly sees that the local mobility is spatially different, with
the occurrence of varying degrees of localization (or even the absence
of it). Thus, this implies that even though all these confined samples
have the same age, the local response to an imposed stress differs
from one to the other, depending upon the initial state. Within an
ensemble, the varying degree of fluidization indicates a distribution
of time-scales for complete fluidization of each initial state; and this
variation is macroscopically captured within $\delta\gamma_{\mathrm
w}(t)$. The increasing peak height of $\delta\gamma_{\mathrm w}(t)$
with decreasing $\sigma_0$ (Fig.~\ref{fig1}b) implies that this
distribution becomes increasingly broad, which we can also see in the
following way. For each $\sigma_0$, we calculate the distribution of
time-scale at which a strain of $0.3$ is reached for the set of $m$
trajectories; the normalised variance ($\Gamma$) of this distribution
is seen to increase with decreasing $\sigma_0$ (see Fig.~\ref{fig1}(d)
for the data corresponding to the ensemble with $t_{\mathrm{age}}=10^4$).

Finally, we investigate the local dynamics during the power-law
creep for $\sigma_0=0.080$ (Fig.~\ref{fig5}).  Note that the local
non-affine motion is extremely slow; even at $t=10790$ (first panel
on the left), only a few faint spots of significant displacements
($\mu_{lm}(t)>0.5\,{d_s}$) are seen.  As time progresses, these spots
expand into larger patches ($t=25730$). By $t=92960$, we see that these
patches form a shear-band-like structure and this remains persistent till
the end of our observation ($t=251905$). In the rest of the system,
even at such long time-scales, the mobility is comparably negligible.
Thus transverse growth of mobilities seems to be impeded here, which
makes the localisation long-lived (unlike when $\sigma > \sigma_d$).

In conclusion, using numerical simulations, we have demonstrated the
existence of spatially heterogeneous dynamics during the onset
of flow under imposed stress. The degree of heterogeneity 
and its lifetime increases with decreasing stress;  and therefore 
creep flow (characterized by a power-law dependence of strain on time)
is associated with long-lived shear-bands.  It is to be noted that in most
studies till now \cite{jlal,schall,besseling,coussot,varnik,pinaki2},
shear-banding has been studied during steady flow, when an external shear-rate
is imposed. In contrast, we use stress as a control parameter and thus
we are able to investigate flow regimes both above and below yield stress.  
There, we see the existence of such dynamical heterogeneities and their
increasing lifetimes. Further, by studying this transient flow \cite{cates} 
from a quiescent glass, we can also check the role of 
aging (the history of which would be lost once the system is in steady flow). 
The evidence of the spatiotemporal heterogeneity that we put forth should
also stimulate the development of theoretical models which
can predict such fluctuations. Till now, most
rheological models \cite{fielding, derec, thomas} predicting creep flows
of various forms do not provide such scenarios. In that aspect, the non-local fluidity model
\cite{bocquet} seems promising; however, the temporal evolution of spatial
patterns of local fluidity, as the material starts flowing, needs to be
worked out. Only recently, a more spatial version of the SGR model
predicted that creep flow is associated with shear-banded velocity
profiles \cite{fielding2} and our simulations confirm that.

Thus, even for a simple planar Couette flow geometry, 
a complex response is observed during the yielding of glasses.
Further studies are necessary to clarify how this transient behaviour
connects to creep observed in uniaxial tensile studies of metallic
\cite{samwer} and polymeric \cite{ediger,depablo,rottler} glasses. For
soft amorphous materials, while more local measurements are necessary
from experiments, simulations can also explore the yielding response
for inhomogeneous stress fields \cite{pois}.

\begin{acknowledgments}
We thank L. Berthier, L. Bocquet, T. Divoux, S. Egelhaaf, W. Kob, M. Laurati, and
T. Voigtmann for useful discussions.  We acknowledge financial support
by the German DFG, project No.~SFB TR 6/A5, and computing time at the
NIC J\"ulich.

\end{acknowledgments}
%

%

%

\begin{thebibliography}{99}
\bibitem{hb} 
The dynamic yield stress ($\sigma_d$) was estimated via constant shear-rate ($\dot{\gamma}_0$) 
simulations  \cite{varnik}.Measured shear stress $\sigma$ is fitted with the Herschel-Bulkley function
$\sigma=\sigma_d[1+(\dot{\gamma_0}/{\gamma_c})^{\eta}]$ to obtain $\sigma_d$.
\end{thebibliography}
\end{document}